\documentstyle[preprint,tighten,aps]{revtex}

\begin{document}
\draft
\preprint{IFUP-TH 8/97}
\title{On the evaluation of universal non-perturbative 
constants in $O(N)$ $\sigma$ models.}
\author{Massimo Campostrini, Andrea Pelissetto, 
Paolo Rossi, and Ettore Vicari}
\address{Dipartimento di Fisica dell'Universit\`a 
and I.N.F.N.,I-56126 Pisa, Italy}

\date{\today}

\maketitle

\begin{abstract}

We investigate the relation between on-shell and zero-momentum 
non-perturbative quantities entering the parametrization of the
two-point Green's function of 
two-dimensional non-linear $O(N)$ $\sigma$ models. 
We present accurate estimates of ratios 
of mass-scales and renormalization 
constants, 
obtained by an analysis of the strong-coupling expansion
of the two-point Green's function.
These ratios allow to connect the exact on-shell results
of Refs.~\cite{Hasenfratz-et-al,Balog-Niedermaier}
with typical zero-momentum lattice evaluations.
Our results are supported by the $1/N$-expansion.

\end{abstract}



{\bf Introduction.}
Physical quantities which are independent of coordinates and 
carry no physical
dimensions, like mass or amplitude ratios, are the best candidates for 
scheme-independent and/or numerical determinations in quantum and
statistical field theories.

In two-dimensional non-linear $O(N)$ $\sigma$ models some
exact results concerning the on-shell 
(large-distance in the Euclidean space) behavior of the two-point 
spin-spin Green's function are known.
Exact formulas have been presented
for the on-shell mass-$\Lambda$-parameter 
ratio~\cite{Hasenfratz-et-al}, and
for the constant $\lambda_1$ of the $O(3)$ $\sigma$
model~\cite{Balog-Niedermaier}, 
which is defined starting 
from a parametrization of the large-momentum asymptotic behaviour 
of the on-shell renormalized two-point correlation function.
No exact off-shell results are known. 
In this letter 
we study the relation between on-shell and zero-momentum 
quantities related to  the two-point Green's function.
This will allow, by using the above-mentioned exact on-shell results,
to determine very accurately (the error being of order $10^{-4}$) 
the asymptotic behavior (i.e. for $\beta\rightarrow \infty$) 
of the non-perturbative quantities which parametrize 
the behavior of the two-point Green's function
at small momentum, such as the second-moment mass and the magnetic
susceptibility.

For the sake of generality, let us discuss the general $O(N)$ model
on a square lattice with nearest-neighbor action
\begin{equation}
S = -N\beta \sum_{x,\mu} \vec{s}(x) \cdot \vec{s}(x+\mu), 
\end{equation}
where $\vec{s}(x)\cdot \vec{s}(x) = 1$.
We have introduced a rescaled inverse-temperature $\beta$, and we
shall use the short-hand 
\begin{equation}
\alpha = {N-2\over 2\pi N\beta}.
\end{equation}
We consider the Fourier transform of the bare 
spin-spin two-point function
\begin{equation} 
G(p;\beta) = \sum_x e^{ipx} \langle \vec{s}(x) \cdot \vec{s}(0) 
\rangle .
\label{eq1}
\end{equation}
In the continuum limit and in the large (euclidean) momentum regime a 
standard one-loop calculation gives 
\begin{equation}
G(p;\beta) \mathop{\longrightarrow}_{p\rightarrow \infty}
2\pi {N-1\over N-2} 
  {\alpha^{\case{N-1}{N-2}}\over p^2}
\overline{\alpha} (p)^{-\case{1}{N-2}},
\label{eq3}
\end{equation}
here $\overline{\alpha} (p)$ denotes the (one-loop) running coupling 
constant
\begin{equation}
 \overline{\alpha} (p) = {\alpha\over 1 + \alpha \ln ap},
\end{equation}
where $a$ is the lattice spacing.
The  corresponding renormalized Green function $G_R(p,M)$  
defined by an on-shell renormalization condition is obtained 
by requiring 
\begin{equation}
G_R(p;M) \mathop{\longrightarrow}_{p^2 + M^2 \to 0}\;\;\;\;
{1\over p^2 + M^2}
\end{equation}
where $M$ is the physical mass-gap. The renormalized correlation function 
can then be parametrized in the large-momentum regime as 
\begin{equation}
G_R(p;M) \mathop{\longrightarrow}_{p\rightarrow \infty}
{\lambda_1(N)\over p^2} 
    \overline{\alpha} (p)^{-{1\over N-2}}.
\label{eq5}
\end{equation}
Notice that Eq.~(\ref{eq5}) is the definition of $\lambda_1(N)$. 
This constant depends crucially on the 
renormalization condition one adopts. 
In turn the bare two-point function (\ref{eq1}) depends 
explicitly on the coupling. In the low-momentum (large-distance)
regime and in the scaling region it can be parametrized by
\begin{equation}
G(p;\beta) \mathop{\longrightarrow}_{p^2 + M^2 \to 0}\;\;\;\;
 {Z(\beta)\over p^2 + M(\beta)^2}.
\label{eq8}
\end{equation}
Renormalization-group arguments lead to the 
following expression for the (universal) asymptotic behavior of 
the mass-gap:
\begin{equation}
M(\beta) =  {1\over a}R(N) \alpha^{-\case{1}{N-2}}
       e^{-\case{1}{\alpha}} \left[ 1 + O(\alpha) \right],
\label{eq9}
\end{equation}
and of   the on-shell renormalization constant
\begin{equation}
Z(\beta) = C(N) \alpha^{\case{N-1}{N-2}} \left[ 1 +
O(\alpha)\right].
\label{eq9b} 
\end{equation}
The constant $R(N)$ is not universal and it can be easily computed
from the exact result of Ref.~\cite{Hasenfratz-et-al} by calculating
the appropriate $\Lambda$-parameter ratio,
which can be obtained by a simple one-loop calculation. For the standard
nearest-neighbor action on the square lattice 
one finds
\begin{equation}
R(N) = \left( {8\over e}\right)^{\case{1}{N-2}}
\Gamma\left( 1 + \case{1}{N-2}\right)^{-1} e^{\case{\pi}{2(N-2)}}
\sqrt{32}.
\label{Hassq}
\end{equation}
The constant $C(N)$ is universal,  
that is independent of the lattice regularization. 
By using Eqs.~(\ref{eq3}), (\ref{eq5}), 
(\ref{eq8}) and (\ref{eq9}), it can be put in relation with $\lambda_1(N)$:
\begin{eqnarray}
\lambda_1(N) = 2\pi {N-1\over N-2} {1\over C(N)}. 
\end{eqnarray}
For $N=3$, using the exact result of Ref.~\cite{Balog-Niedermaier}, i.e. 
\begin{equation}
\lambda_1(3) = {4\over 3\pi^2},
\label{lambda1}
\end{equation}
we get 
\begin{equation} 
 C(3) = 3\pi^3.
\label{c3exa}
\end{equation}

The main purpose of the present letter is to compute the relation
between $R(N)$ and $C(N)$ and the corresponding zero-momentum
quantities, which are much easier to compute in Monte Carlo
lattice simulations. 
We will obtain results which are quite accurate, although not exact.
 
Typical lattice calculations lead to estimates of moments 
of the two-point function:
\begin{equation}
m_{2i} \equiv \sum_x (x^2)^i \langle \vec{s}(0)\cdot\vec{s}(x)\rangle.
\end{equation}
In particular $\chi\equiv m_0$.
If we now parametrize the function $G(p;\beta)$ 
around $p=0$ by
\begin{equation}
G(p;\beta) \approx {Z_G(\beta)\over p^2 + M^2_G(\beta)}
\end{equation}
($M_G$ is by definition the inverse of the so-called 
second-moment correlation length), we obtain the relationships 
\begin{eqnarray}
M^2_G &=& {4 m_0\over m_2}, \label{eq14} \\
Z_G &=& \chi M^2_G = {4 m_0^2\over m_2}. 
\label{eq15}
\end{eqnarray}
Standard renormalization-group arguments lead again to 
\begin{eqnarray}
M_G(\beta) &=&  R_G(N) \alpha^{-\case{1}{N-2}}
       e^{-\case{1}{\alpha}} \left[ 1 + O(\alpha)\right],\\
Z_G(\beta) &=&  C_G(N) \alpha^{\case{N-1}{N-2}} \left[ 1 + O(\alpha)\right].
\end{eqnarray}
$C_G(N)$ and $R_G(N)$ differ from the 
corresponding quantities $C(N)$ and $R(N)$ (except at $N=\infty$ where
the theory is Gaussian).
In order to calculate for these 
two quantities, we will thus investigate the dimensionless ratios
\begin{eqnarray}
S_M  =  \lim_{\alpha\to 0} {M^2\over M_G^2},\quad\quad
S_Z =  \lim_{\alpha\to 0} {Z^{-1}\over Z_G^{-1}}. 
\end{eqnarray}
These quantities are not exactly known and we will provide here 
rather accurate strong-coupling estimates, supported by
a $1/N$ analysis.

\medskip
{\bf Strong-coupling estimates of dimensionless RG invariant quantities.}
As shown in Ref.~\cite{ON-d2-a}, 
strong-coupling analysis may provide quite accurate continuum-limit
estimates when applied directly to dimensionless
renormalization-group invariant ratios of physical quantities.  
The basic idea is that
any dimensionless renormalization-group invariant quantity
$R(\beta)$ behaves, for sufficiently large $\beta$, as
\begin{equation}
R(\beta)-R^*\sim M(\beta)^2,
\label{scalR}
\end{equation}
where $R^*$ is its fixed point (continuum) value and $M(\beta)$ goes
to zero for $\beta\rightarrow \infty$. 
Hence a reasonable estimate of $R^*$
may be obtained at the values of $\beta$ corresponding to large but
finite correlation lengths, where the function $R(\beta)$ flattens.
This is essentially the same idea underlying Monte Carlo studies of
asymptotically free theories, based on the identification of the
so-called scaling region.
Strong-coupling estimates of physical quantities 
may be obtained by evaluating  approximants of their
strong-coupling series 
at values of $\beta$ corresponding to reasonably large correlation
lengths, e.g. $\xi\gtrsim 10$. 
Scaling is then checked by observing the stability
of the results varying $\beta$. 

In a strong-coupling analysis it is crucial to
search for improved estimators of the quantities at hand,
because better estimators can greatly improve the stability of the
extrapolation to the critical point. 
Our search for
optimal estimators was guided by the large-$N$ limit of 
lattice $O(N)$ $\sigma$ models which is a Gaussian theory.
We chose estimators which are perfect for $N=\infty$, i.e. do not
present off-critical corrections to their critical value.

On the lattice, in the absence of a strict rotation invariance, 
one may define different estimators of the mass-gap 
$M$ having the same critical
limit.  On the square lattice one may consider $\mu$ obtained by the
long-distance behavior of the side wall-wall correlation constructed
with $G(x)$, or equivalently the solution of the equation
$G^{-1}(i\mu,0;\beta)=0$.
At a finite order $q$ of the strong-coupling expansion, 
the wall-wall spin-spin correlation function $G_w(z)$ at a 
distance larger than $q/3$ exponentiates exactly, i.e. 
for $|z|>q/3$ it can be written as 
\begin{eqnarray}
G_w(z;\beta) = A(\beta) e^{- \mu(\beta) |z|}.
\end{eqnarray}
In the context of a strong-coupling
analysis, it is convenient to use another estimator of the mass-gap
derived from $\mu(\beta)$~\cite{ON-d2-a}:
\begin{equation}
M_{\rm s}^2(\beta) = 2\left( {\rm cosh} \mu(\beta) - 1\right).
\label{MM}
\end{equation}
Moreover, by comparison with Gaussian model, 
we consider the following estimator of $Z$
\begin{eqnarray}
Z_s(\beta) = 2 A(\beta) \sinh \mu(\beta).
\end{eqnarray}
In practice, when the strong-coupling expansion of 
$G(x;\beta)$ is known to order $q$, $M_s^2(\beta)$ can be determined up to 
about $2 q/3$ orders~\cite{SCchiral}, 
and the same precision is therefore achieved in the determination 
of $Z_s(\beta)$. 
Estimators of the zero-momentum mass $M_G$ and renormalization
constant $Z_G$ can be easily extracted 
from Eqs.~(\ref{eq14}) and (\ref{eq15}).  
So in order to estimate $S_M$ and $S_Z$ one should study the continuum
limit of the ratios 
\begin{equation}
\overline{S}_M(\beta)\equiv {M_s^2\over M_G^2}, \quad\quad
\overline{S}_Z(\beta)\equiv {Z_G\over Z_s}.
\end{equation} 
In the large-$N$ limit $\overline{S}_M(\beta)=\overline{S}_Z(\beta)=1$ 
independently of $\beta$.
Quantities having the same properties of $M_s^2$ and $Z_s$ can be
conceived also on the honeycomb and triangular
lattices~\cite{ON-d2-a}, thus leading to analogous 
definitions of $\overline{S}_M(\beta)$ and $\overline{S}_Z(\beta)$.

\medskip
{\bf Analysis of the strong-coupling series.}
We have analyzed the strong-coupling series of
$\overline{S}_M$ and $\overline{S}_Z$ on the square lattice, where 
the available series~\cite{ON-d2-a} are of the form 
$1+\beta^6\sum_{i=0}^{10} a_i\beta^i$ in both cases. 
An analogous analysis has been performed on the honeycomb and triangular
lattices within their nearest-neighbor formulations
using the available series of the two-point function~\cite{ON-d2-a}.
The analysis of strong-coupling series calculated 
on different lattices offers a possibility of testing
universality. On the other side, once universality is assumed, it
represents a further check for possible systematic errors, whose
estimate is usually a difficult task
in strong-coupling extrapolation methods such as those based on Pad\'e
approximants and their generalizations.

In Table~\ref{detailssq} we present some details of the results 
obtained on the square lattice.
There we report  estimates of $\overline{S}_M$ and 
$\overline{S}_Z$ at various values of $\beta$, where
the correlation length is reasonably large.
Such estimates of $\overline{S}_M$ and $\overline{S}_Z$
have been  obtained by resumming the strong-coupling series of
$\left( \overline{S}_M-1\right)/\beta^6$ and 
$\left( \overline{S}_Z-1\right)/\beta^6$ 
by Dlog-Pad\'e approximants (DPA's), in which  the standard
Pad\'e resummation is applied to the series of the
logarithmic derivative, and then the original quantity
is reconstructed. This method of resummation turned out to give
the most stable results (we also tried simple Pad\'e approximants
and first order integral approximants).
For a $n$th order series, 
we considered $[l/m]$ DPA's having
\begin{equation}
l+m+1\geq n-2, \quad\quad l,m\geq  \case{n}{2} - 2.
\end{equation}
As estimate at a given $\bar{\beta}$ 
we took the average of the values of the   
non-defective approximants constructed using all available
terms of the series. As an indicative error 
we considered
the square root of the variance around the estimate of the results 
from all non-defective approximants specified above.
This quantity should give an idea of the spread of the results from different
approximants. 
Approximants are considered defective when
they present spurious singularities close to the real axis for ${\rm
Re} \beta \lesssim \bar{\beta}$.

The precision of the results is satisfactory even for values of
$\beta$ where the correlation length is quite large. Furthermore
scaling is well verified. 
Then, assuming scaling, we extracted estimates of the 
corresponding continuum limit, which are reported in
Table~\ref{summary}.
There we report results obtained on the square, 
honeycomb and triangular lattices, and  for several values of $N$
($N=3,8,16$, where the last two large values of $N$ has been 
considered in order
to make a comparison with the large-$N$ analysis, see later).
Errors represent a rough  estimate of the uncertainty,
which is quite small.
Universality among different lattice formulations is well verified. 
Our final estimates for $N=3$ are
\begin{equation}
S_M = 0.9987(2) ,\quad\quad S_Z = 1.0025(4).
\label{N3res}
\end{equation}

There are some estimates of $S_M$ obtained by high-statistics Monte
Carlo simulations that are worth being mentioned for comparison. 
Monte Carlo simulations at $N=3$~\cite{Meyer}
gave $S_M=0.9988(16)$ at $\beta={1.7/3}=0.5666...$ ($\xi\simeq 35$), 
and $S_M=0.9982(18)$ at $\beta=0.6$ ($\xi\simeq 65$),
leading to the estimate $S_M=0.9985(12)$.
From the data of Ref.~\cite{tomeu} one derives 
$S_M=0.996(2)$ for $N=3$ and $S_M=0.9978(8)$ for $N=8$.
These numbers compare well with our strong-coupling calculations,
which appear to be much more precise.

\medskip
{\bf The two-point Green's function at small momentum.}
The fact that both $S_M\approx1$ and $S_Z\approx1$ should not come as 
a surprise. Indeed it was shown in Ref.~\cite{ON-d2-a} that in the region 
$p^2 \lesssim M^2_G$ the spin-spin two-point function is essentially 
Gaussian with very small corrections. 
The inverse two-point function can be studied around $p^2=0$ by
expanding it in powers of $p^2$:
\begin{equation}
G^{-1}(p) = Z^{-1}_G M^2_G
\left[ 1 + {p^2\over M^2_G} + 
      \sum_{i=2}^\infty  c_i \left( {p^2\over M^2_G}\right)^i\right].
\end{equation}
Analysis based on various approaches have shown 
that in two- and three-dimensional
$O(N)$ models the following relations hold~\cite{ON-d2-a,ON-d2-b,ON-d3}
\begin{equation}
c_2\ll 1, \quad\quad c_i \ll c_2 \quad {\rm for} \quad i>2. 
\label{crel}
\end{equation}
Then neglecting 
all $c_i$, $i\ge 3$ and terms of order $c_2^2$ one may write
the following expression for the inverse two-point function
\begin{equation}
G^{-1}(p)  \,{\approx} \, Z^{-1}_G M^2_G
\left[ 1 + c_2 + {p^2\over M^2_G} \right] 
\left[ 1 - c_2 + c_2 {p^2\over M^2_G} \right], 
\end{equation}
which should give a good approximation of the two-point
Green's function in the region $|p^2|\lesssim M_G^2$.
 As a consequence one obtains the following approximate relations
\begin{eqnarray}
S_M &\simeq& 1 + c_2, \label{smapp} \\ 
S_Z &\simeq& 1 - 2 c_2.\label{szapp}
\label{eq23}
\end{eqnarray}

In order to have a check of Eqs.~(\ref{smapp}) and (\ref{szapp}),
we have estimated $c_2$ by a strong-coupling analysis.
Again guided by the large-$N$ limit,
we considered the following estimator of $c_2$
\begin{equation}
\overline{c}_2(\beta) = 1 - {m_4\over  64 m_0} M_G^4+ {1\over 16}M_G^2.
\end{equation}
In the continuum limit $\overline{c}_2(\beta)\rightarrow c_2$.
In the large-$N$ limit $\overline{c}_2(\beta)=c_2=0$ independently of
$\beta$ for all square, honeycomb and triangular lattices.
We mention that on the square lattice,
where $G(x)$ has been calculated up to 21st order~\cite{ON-d2-a},
 the available series of 
$\overline{c}_2$ is of the form $\beta^4\sum_{i=0}^{15} a_i\beta^i$.
Details of the analysis of $\overline{c}_2$ on the square lattice can be found
in Table~\ref{detailssq}. Final estimates are reported in
Table~\ref{summary}, where also results obtained on the honeycomb and
triangular lattices are reported.
Numerically in the case of $O(3)$ we obtained $c_2 \simeq -0.0012$,
which nicely confirms Eqs.~(\ref{smapp}) and (\ref{szapp}).

\medskip
{\bf $1/N$ expansion.}
We have also evaluated $S_M$ and $S_Z$ in the context of the 
$1/N$ expansion, which 
was found to be a fairly accurate approach to the evaluation 
of amplitude ratios in two-dimensional $O(N)$ models for $N>2$.
An analytic computation of the two-point function in the region 
$p^2 + M^2 \approx 0$ leads to the following result:
\begin{equation}
G^{-1}(p;\beta) \approx {p^2 + M^2\over 2 \pi \alpha} 
\left[ 1 + {1\over N} \left(\log {4\over \pi \alpha} - \gamma_E - 3 \right) 
  + O(1/N^2)\right],
\end{equation}
and therefore
\begin{equation}
\lambda_1(N) = 1 + {1\over N}\left( \log {4\over \pi} + \gamma_E - 2\right) + 
  O(1/N^2).
\end{equation}
By evaluating the $O(1/N)$ contribution to the 
on-shell-renormalized self-energy in the region around 
$p=0$, one finds
\begin{eqnarray} 
S_M &=& 1 - {0.00645105\over N} + O\left( {1\over N^2}\right), \\
S_Z &=& 1 + {0.01317046\over N} + O\left( {1\over N^2}\right) ,\\
c_2 &=& - {0.00619816\over N} + O\left( {1\over N^2} \right) ,\\
c_3 &=&  {0.00023845\over N} + O\left( {1\over N^2} \right) ,\\
c_4 &=& - {0.00001344\over N} + O\left( {1\over N^2} \right) ,
\end{eqnarray}
etc...
The coefficients of the $O(1/N)$ terms show consistency with
Eqs.~(\ref{smapp}) and (\ref{szapp}) 
within errors of order $10^{-4}$.
Inclusion of $c_3 \approx 2.4 \cdot 10^{-4}/N$ would squeeze the error 
to $O(10^{-5})$.
Strong-coupling estimates reported in Table~\ref{summary}
clearly approach the large-$N$ asymptotic
regime predicted by the above equations. 
In particular quantitative agreement
(within the uncertainty of our strong-coupling calculations)
is found at $N=16$. This represents a further check
of the analysis employed in order to get strong-coupling
estimates in the continuum limit. 

It is worth mentioning that $c_2$, $S_M$ and $S_Z$ have been
also calculated to $O(\epsilon^3)$ 
within the $\phi^4$ formulation of $O(N)$ models
in $4-\epsilon$ dimensions~\cite{ON-d3}.
The approximate relations (\ref{smapp}) and (\ref{szapp}) are confirmed
even in the $\epsilon$-expansion, whose validity should not be
related to the specific value of $N$.
Furthermore a semi-quantitative comparison,
inserting the value $\epsilon=2$ in the $\epsilon$-expansion
formulae, provides the correct order of magnititude.

\medskip
{\bf Conclusions.}
Putting together the exact formulas
(\ref{Hassq}) and (\ref{c3exa}) and our strong-coupling estimates
of the ratios $S_M$ and $S_Z$, we arrive at the following 
results
\begin{eqnarray}
R_G(3) &=& {R(3)\over \sqrt{S_M}}=80.139(8),\label{rg3}\\
C_G(3) &=& C(3)\times S_Z= 93.25(3).
\label{cg3}
\end{eqnarray}

For comparison we mention the existing Monte Carlo results
concerning the ratio $C_G(3)/R_G(3)^2$.
Ref.~\cite{CEMPS} quotes $C_G(3)/R_G(3)^2= 0.0146(11)$,
which has been obtained 
by employing finite-size-scaling based techniques allowing
to reach correlation lengths up to $O(10^5)$.
Ref.~\cite{tomeu} quotes 
$C_G(3)/R_G(3)^2= 0.0138(2)$, which has been
obtained by standard Monte Carlo simulations
up to $\xi\simeq 130$,  and where the error is just statistical.
An attempt to estimate
the systematic error due to violations of asymptotic scaling
would give the number $0.0138(2)(7)$\footnote{
In Ref.~\cite{tomeu} the estimate of the ratio 
$C_G(3)/R_G(3)^2$ has been obtained by using the Symanzik
tree-improved action and considering the so-called energy scheme.
In order to estimate the systematic error due to 
violations of asymptotic scaling, we have considered the difference
between the results obtained by using two-loop and three-loop
formulas in the fit of Monta Carlo data.
Two-loop and three-loop formulas lead 
to $0.0145(2)$ and $0.0138(2)$ respectively.},
  where the second number within brackets
is the systematic error estimated by us.
These numbers compare very well with our corresponding estimate
derived from Eqs.~(\ref{rg3}) and (\ref{cg3}),
i.e.  $C_G(3)/R_G(3)^2= 0.01452(5)$.


\newpage

\begin{table}
\caption{
Estimates of $\overline{S}_M$, $\overline{S}_Z$ 
and $\overline{c}_2$ obtained from DPA's
of the strong-coupling series on the square lattice
evaluated at various values of $\beta$ where
the correlation length $\xi$ is reasonably large.
For example at $N=3$:  
$\xi(\beta=0.45)\simeq 8$, 
$\xi(\beta=0.50)\simeq 11$, 
$\xi(\beta=0.55)\simeq 25$, and 
$\xi(\beta=0.60)\simeq 65$; at $N=8$:
$\xi(\beta=0.50)\simeq 5$, 
$\xi(\beta=0.55)\simeq 8$, and 
$\xi(\beta=0.60)\simeq 12$.
\label{detailssq}}
\begin{tabular}{ccr@{}lr@{}lr@{}lr@{}lr@{}l}
\multicolumn{1}{c}{$N$}&
\multicolumn{1}{c}{}&
\multicolumn{2}{c}{$\beta=0.45$}&
\multicolumn{2}{c}{$\beta=0.50$}&
\multicolumn{2}{c}{$\beta=0.55$}&
\multicolumn{2}{c}{$\beta=0.60$}\\
\tableline \hline
3 &$(\overline{S}_M-1)\times 10^3$&$-$1&.01(3) &$-$1&.2(1) 
  &$-$1&.1(2) &$-$0&.9(3) \\  
  &$(\overline{S}_Z-1)\times 10^3$&   2&.04(6) 
  &2&.4(2) &   2&.2(4) &   1&.4(7) \\  
  &$\overline{c}_2\times 10^3$    
  &$-$1&.16(4) &$-$1&.3(2) &$-$1&.4(3) &$-$1&.3(4) \\  
\hline
8  &$(\overline{S}_M-1)\times 10^3$&$-$0&.54(1) 
  &$-$0&.59(2) &$-$0&.57(5) &$-$0&.5(1) \\  
  &$(\overline{S}_Z-1)\times 10^3$&   1&.11(3) &   1&.2(1)  
  &   1&.2(1)  &   1&.1(2) \\  
  &$\overline{c}_2\times 10^3$    
&$-$0&.58(2) &$-$0&.63(5) &$-$0&.6(1)  &$-$0&.6(2) \\  
\hline
16&$(\overline{S}_M-1)\times 10^3$&$-$0&.32(4) &$-$0&.35(6) 
  &$-$0&.4(1)  &$-$0&.4(2) \\  
  &$(\overline{S}_Z-1)\times 10^3$&   0&.7(1)  &   0&.7(1)  
  &   0&.8(2)  &   0&.9(4) \\  
  &$\overline{c}_2\times 10^3$    
  &$-$0&.31(1) &$-$0&.34(3) &$-$0&.35(6) &$-$0&.35(10) \\
\end{tabular}
\end{table}

\begin{table}
\caption{
We report estimates of $S_M$, $S_Z$ and $c_2$ from
the strong-coupling expansion on the
square, honeycomb and triangular lattice,
and $1/N$ expansion of the continuum formulation
of the non-linear $O(N)$ $\sigma$ model.
Final strong-coupling estimates are taken at
$\beta$-values corresponding to $\xi\simeq 10$.
\label{summary}}
\begin{tabular}{ccr@{}lr@{}lr@{}lr@{}l}
\multicolumn{1}{c}{$N$}&
\multicolumn{1}{c}{}&
\multicolumn{2}{c}{$S_M$}&
\multicolumn{2}{c}{$S_Z$}&
\multicolumn{2}{c}{$c_2$}\\
\tableline \hline
3 & square & 0&.9988(2)  & 1&.0024(4) & $-$1&.3(2)$\times 10^{-3}$  \\
  & honeycomb & 0&.9986(3) & 1&.0027(4) & $-$1&.2(2)$\times 10^{-3}$  \\
  & triangular &  0&.9985(5) & 1&.003(1) & $-$1&.2(3)$\times 10^{-3}$\\
  & $O(1/N)$  & 0&.9978  & 1&.0044 & $-$2&.07$\times 10^{-3}$  \\\hline
8 & square & 0&.99943(5) & 1&.0012(1) & $-$0&.6(1)$\times 10^{-3}$  \\
  & honeycomb & 0&.9994(1)  & 1&.0011(2)  & $-$0&.7(1)$\times 10^{-3}$  \\
  & $O(1/N)$  & 0&.99919  & 1&.00164 & $-$0&.77$\times 10^{-3}$  \\\hline
16 & square & 0&.9996(1)  & 1&.0008(2) & $-$0&.35(5)$\times 10^{-3}$  \\
   & honeycomb & 0&.9997(1)  & 1&.0006(1) & $-$0&.36(3)$\times 10^{-3}$  \\
  & $O(1/N)$  & 0&.99960 & 1&.00082 & $-$0&.39$\times 10^{-3}$  \\
\end{tabular}
\end{table}

\end{document}